\documentclass[aps,prl,twocolumn,superscriptaddress,amsmath,showpacs]{revtex4-1}
\usepackage{graphicx}
\usepackage{subfigure}

\usepackage[T1]{fontenc}
\usepackage[latin1]{inputenc}
\usepackage[rgb,svgnames]{xcolor}

\begin{document}
\newcommand{\ket}[1]
{\left|#1\right\rangle}

\newcommand{\bra}[1]
{\left\langle #1\right|}

\newcommand{\braket}[2]
{\left\langle #1\right|\left.#2\right\rangle}

\title{Heralded generation of a micro-macro entangled state}

\author{Ulrik L. Andersen and Jonas S. Neergaard-Nielsen}
\affiliation{Department of Physics, Technical University of Denmark, Fysikvej, 2800 Kgs. Lyngby, Denmark}

\date{\today}

\begin{abstract}
Using different optical setups based on squeezed state and photon subtraction we show how optical entanglement between a macroscopic and a microscopic state - the so-called Schr\"{o}dinger cat state or micro-macro state - can be generated.   
The entangled state is heralded and is thus produced a priori in contrast to previous proposals. 
We define the macroscopicity of the macroscopic part of the state as their mean distance in phase space and the success rate in discriminating them with homodyne detection, and subsequently, based on these measures we investigate the macroscopicity of different states. 
Furthermore, we show that the state can be used to map a microscopic qubit onto a macroscopic one thereby linking a qubit processor with a qumode processor.
\end{abstract}

%\pacs{}

\maketitle

\section{Introduction}
Quantum superpositions are at the heart of quantum mechanics. Simple examples are two-dimensional superpositions of microscopic systems such as two-level atoms, the polarization of a single photon and the spin of an electron. Being at a microscopic level, these superpositions are readily accepted but if they are brought to the macroscopic level they become counter-intuitive and hardly imaginable. This is commonly illustrated by the famous Gedankenexperiment of Schr\"{o}dinger in 1935 where he considers the superposition of a cat in two distinct states: dead and alive~\cite{sch}. In this experiment the cat is entangled with a microscopic degree of freedom, namely the discrete energy levels of an atom. Therefore, the proposal does not only demonstrate the superposition principle on a macroscopic scale but also the peculiar feature of nonlocality~\cite{einstein}. 

In recent years there has been a strong focus on bringing quantum mechanics into a macroscopic realm through careful state engineering and suppression of environmental noisy modes~\cite{leggett}. Macroscopic superpositions of atomic clouds~\cite{polzik}, superconducting circuits~\cite{wal,friedman}, ions~\cite{leib} and microwaves~\cite{raimond} have been prepared, and there are proposals on how to push this into a regime of massive systems~\cite{penrose} and even living organisms~\cite{isart}. 

In the pure optical regime there have also been a number of successful attempts to generate macroscopic quantum states. One example is the generation of coherent state superpositions by means of photon subtraction of a squeezed vacuum state \cite{wen,our,neer,wak,NIST}. Although being useful for quantum information processing~\cite{lund}, strictly speaking, these states are not cat states in the spirit of Schr\"{o}dinger as the macroscopic states (here coherent states) are not entangled with a microscopic degree of freedom. Another realization of a macroscopic state is the so-called micro-macro state in which the polarization degree of freedom of a single photon is entangled with distinct states containing a large number of photons~\cite{martini2}. These states have been produced in a non-heralded fashion~\cite{lombardi,martini} and their characterization has been discussed in several papers~\cite{martini,sekat,spag,simon}.
 
In this paper we suggest a number of different strategies for generating a heralded micro-macro state based on standard quantum optical tools. As opposed to previous proposals and experiments on micro-macro entanglement, in the present scheme the entanglement is produced between a microscopic photon number (or phase) degree of freedom and a macroscopic wave degree of freedom also known as a qumode~\cite{aki,loock}. In some of the suggested realizations the macroscopic states are classically distinguishable. This distinguishability is referred to as the macroscopicity of the state which will be discussed in relation to different measures; the mean phase space distance between the two macroscopic states and their distinguishability with respect to a quadrature measurement. Finally, we also discuss how these states can be used to map a microscopic qubit onto a macroscopic qumode by means of teleportation.

\section{Squeezing-induced micro-macro state}

A micro-macro state can be written as 
\begin{eqnarray}
|m_+\rangle|\Phi_+\rangle+|m_-\rangle|\Phi_-\rangle
\label{micromacro}
\end{eqnarray}
where $\{|m_+\rangle,|m_-\rangle\}$ are orthogonal microscopic states and $|\Phi_+\rangle,|\Phi_-\rangle$ are states that are macroscopic in a certain degree of freedom. The micro-macro degrees of freedom are often considered to be the photon numbers but it could also be another observable such as the quadrature observable. Moreover, one could consider different observables associated with the microscopic and the macroscopic states (as will be used later in this paper). Before discussing the definition of macroscopicity, we will consider some particular realizations of the heralded micro-macro state.      

In the experiment of De Martini \textsl{et al.} \cite{martini}, the polarization degree of freedom of a single photon was entangled with the polarization degree of freedom of a macroscopic state containing a large number of photons exceeding $10^4$. This was enabled by unitary amplification (using a phase-insensitive, polarization nondegenerate two-mode squeezer) on one half of a polarization entangled photon pair produced by parametric downconversion. The experiment was carried out in the coincidence basis, and thus the resulting micro-macro entangled state was generated a posteriori: Although the amplification process was deterministic, the generation of polarization entangled photons was non-heralded. Heralded generation of polarization entangled photons has recently been realized~\cite{pan,zeilinger} in complicated setups and its extension to produce a heralded micro-macro state renders the setup even more challenging.   

% ---------------------------------------------------------
%
\begin{figure}[t]
\begin{center}
\includegraphics[width=0.40\textwidth]{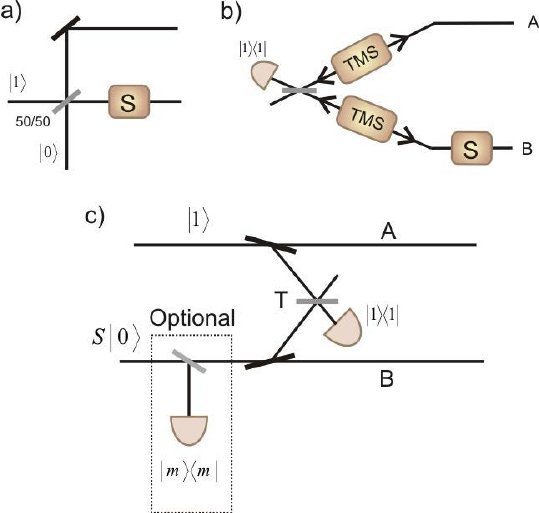}
\caption{(Color online) Schematic setup of the proposed schemes. $S$ denotes a single mode squeezing operation and TMS is a two-mode vacuum squeezer.}
\label{fig:setups}

\end{center}
\end{figure} 
%
% ---------------------------------------------------------

Instead of amplifying polarization entangled photons with a two-mode squeezer, we suggest to amplify a path-entangled single photon with a single-mode squeezer. The setup is shown in Fig.~\ref{fig:setups}(a). A single photon is prepared (e.g. by heralded parametric down conversion) and subsequently split on a balanced beamsplitter to generate a path-entangled single-photon state. One mode is then amplified using a single-mode squeezer (similar to a phase-sensitive amplifier) to produce the following state
\begin{eqnarray}      
|\Phi\rangle=\frac{1}{\sqrt{2}}\left(|1\rangle |\Phi_+\rangle+|0\rangle |\Phi_-\rangle\right)
\label{cat}
\end{eqnarray}
where $|\Phi_+\rangle=S(r)|0\rangle$ and $|\Phi_-\rangle=S(r)|1\rangle$ are orthogonal states
with an average photon number of $\langle\Phi_+|n|\Phi_+\rangle=\sinh r$ and $\langle\Phi_-|n|\Phi_-\rangle=2\sinh r+\cosh r$, respectively. $S(r)=\exp(\frac{r}{2}(a^{\dagger 2}-a^2))$ is the squeezing operator, $r$ is the squeezing parameter, $a$ is the annihilation operator and $|0\rangle$($|1\rangle$) represent the vacuum (single-photon) state. The number of photons of the squeezed states is limited by the degree of attainable squeezing. For strongly pumped optical parametric amplifiers, however, this number can be very large as shown in the experiment of Ref.~\cite{martini}. In this regime the states become macroscopic in the photon number degree of freedom. 

Alternatively, one could substitute the squeezing operation with a much simpler displacement operation (represented by $D(\beta)$ where $\beta$ is the displacement amplitude) in which the macroscopic state would be $|\Phi_+\rangle=D(\beta)|0\rangle=|\beta\rangle$ and $|\Phi_-\rangle=D(\beta)|1\rangle$, that is, a coherent state and a displaced single photon state. This scheme is related to the one suggested in Ref.~\cite{se}. 
However, the two macroscopic states cannot be made perfectly macroscopically distinguishable in their photon numbers. This will require a microscopic parity measurement.   

\section{Micro-macro entanglement and remote preparation}

Since the macroscopic states $|\Phi_\pm\rangle$ are orthogonal, the micro-macro state in (\ref{cat}) is maximally entangled. However, for remote preparation of the entangled state, that is, preparing the Fock state components at one site (Alice) and the squeezed state components at another site (Bob), one of the modes must be sent through a lossy channel which inevitably will lead to a degradation of the entanglement rendering the state non-maximally entangled. It is, however, possible to circumvent the propagation losses as the delocalized single photon can be heralded at a distance using the method outlined in Ref. \cite{brask,grangier,zoller}: The generation of the path-entangled photon state can be implemented employing two sources of two-mode squeezed states (one at Alice (A) and one at Bob (B)). One mode from each source combines at a symmetric beamsplitter, and the measurement of a single photon heralds the desired state. The remotely prepared path-entangled single-photon state is then subsequently squeezed at one site (e.g. at Bob) to generate the required state (see Fig. 1(b)). Using this approach, maximally entangled micro-macro states can be generated at a distance independent on the losses between the two sites. We note, however, that the increase in the state purity is traded for a decrease in the generation rate. 

Characterizing the entanglement of the micro-macro state has been debated in the literature. In the experiment of Ref.~\cite{martini}, the entanglement was quantified by using a Stokes parameter measurement to measure the polarization degree of freedom of the single photon and a special filter detector to discriminate the two multi-photon states~\cite{simon}. Homodyne tomography could not be used in this experiment to fully characterize the state due to the a-posteriori type of generation scheme. On the contrary, the schemes suggested in this paper are based on heralding (that is, the states are not produced a posteriori), and thus homodyne detection can be used to fully characterize the state. With two-mode homodyne tomography, the full density matrix can be reconstructed and the entanglement can be evaluated~\cite{lvov}. An alternative to homodyne tomography is to unsqueeze the macroscopic states and subsequently use a photon counter to measure the resulting Fock states~\cite{simon}.  

\section{Micro-macro entanglement via single-photon subtraction}

The squeezing of a single photon as introduced above is identical to subtracting a single photon from a squeezed vacuum state. This leaves open another way of preparing the entangled state in (\ref{cat}) at a distance. The circuit is illustrated in Fig. \ref{fig:setups}(c) (without the optional box). The idea is to jointly subtract a single photon from two locally prepared quantum states: a single-photon state at Alice's site and a squeezed vacuum state at Bob's site. The joint subtraction is enabled by three beamsplitters and a single-photon counter which is described by the non-unitary operation $\sqrt{T}a_A+\sqrt{1-T}a_B$ in the limit of very low reflection of the tapping beamsplitters. $a_A$ and $a_B$ are the annihilation operators acting on the modes of Alice and Bob, respectively, and $T$ is the transmission coefficient of the "measurement" beamsplitter (see Fig. \ref{fig:setups}(c)). The transformation reads
\begin{eqnarray}     
\label{jointsub}
&&\ket{1} S(r) \ket{0}  \rightarrow  \ket{\Phi} \\ 
&=& (\sqrt{T} a_A + \sqrt{1-T} a_B) \ket{1} S(r) \ket{0}  \nonumber  \\
&=& \sqrt{T} \ket{0} S(r) \ket{0} + \sqrt{1-T} \ket{1} a_B S(r) \ket{0} \nonumber \\
&=& \ket{0} \left( \sqrt{T} S(r) \ket{0}\right) + \ket{1} \left( \sqrt{1-T} \sinh r S(r) \ket{1}\right) \nonumber
\end{eqnarray}
which is identical to (\ref{cat}) for $\sqrt{T}=\sqrt{1-T} \sinh r$ (up to a bit-flip operation), and therefore the schemes in Figs. 1(a) and 1(c) are identical assuming perfect photon subtraction and $T=\sinh^2 r/(1+\sinh^2 r)$. Note that $\ket{\Phi}$ has not been normalized -- the photon subtraction is probabilistic and therefore does not preserve the normalization.

% ---------------------------------------------------------
%
\begin{figure*}[t]
\begin{center}
\includegraphics[width=.9\textwidth]{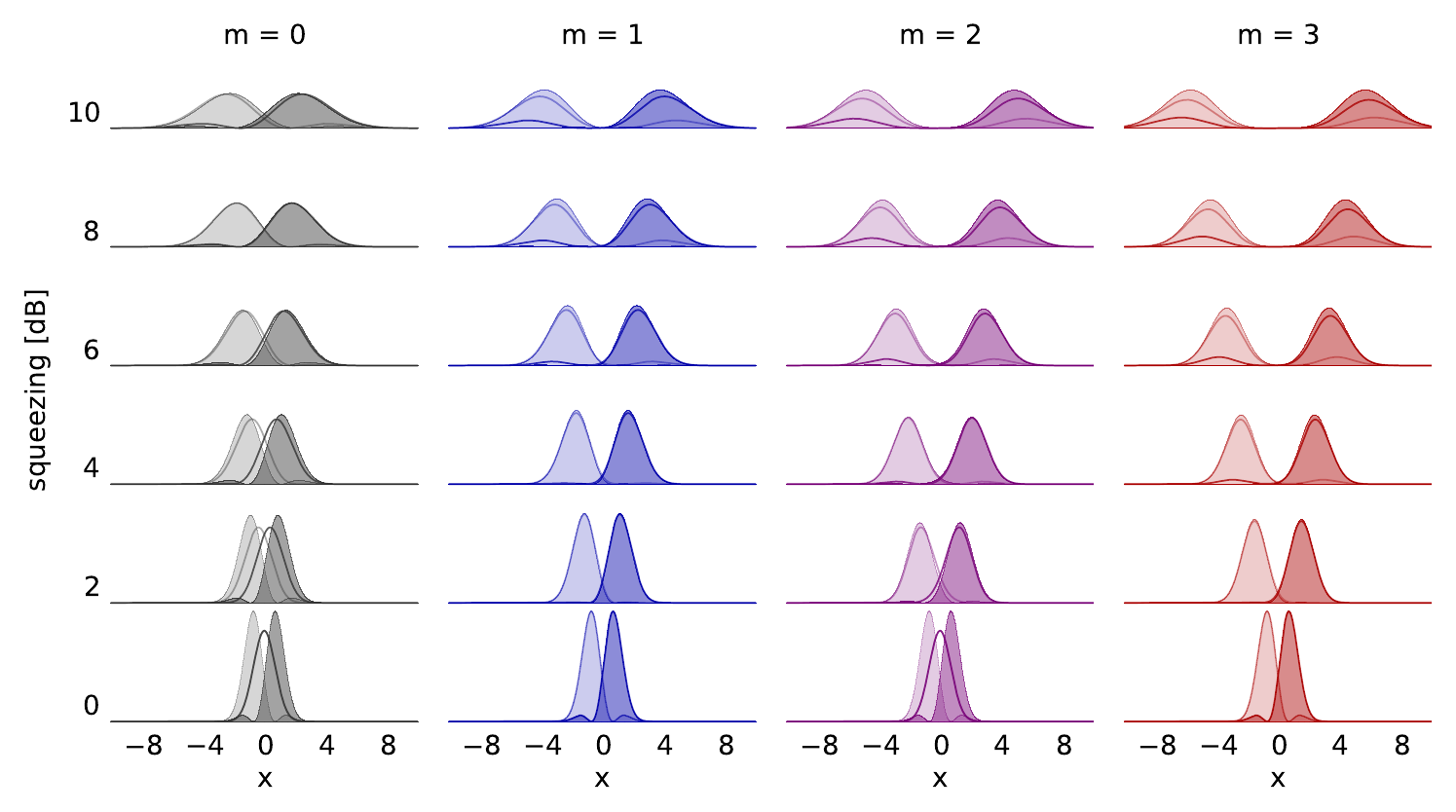}
\caption{(Color online) Quadrature probability distributions for the macroscopic components $\ket{\Psi_+}$ (dark) and $\ket{\Psi_-}$ (light) generated using the setup in Fig.\ref{fig:setups}(c) for different photon subtractions, $m$, and different degrees of squeezing. The solid line wavepackets correspond to a transmission coefficient of $T=1/2$ of the beamsplitter for joint subtraction whereas the shaded packets are associated with the balanced case where $T=T_{bal}$.}
\label{fig:quad}

\end{center}
\end{figure*}   
%
% ---------------------------------------------------------

As the purity of the resulting state is independent on the losses of the joint measurement and the channel, the state can be prepared remotely without degradation. However, as pointed out above, the preparation rate will depend on the losses. The generation strategy (in Fig. \ref{fig:setups}(c)) has the additional practical advantage of using only off-line non-classical transformations which means that there is no need of injecting the non-classical state into a non-linear element as is the case in Fig. \ref{fig:setups}(a) and \ref{fig:setups}(b). Alternatively, an off-line squeezing operation can be implemented using homodyne-based electro-optical feed-forward \cite{filip,huck}.

\section{Macroscopicity of the micro-macro state in phase space}

To qualify as a cat state, the two macroscopic states should be detectable with a coarse-grained detector, that is, the measurement outcomes should be well separated in a particular degree of freedom. For example, they should have macroscopically different photon numbers or macroscopically different quadrature values such that a coarse-grained (or noisy) intensity or homodyne detector can easily discriminate the two states. We refer to this macroscopic distinguishability as the mascroscopicity of the components. We note that according to this definition, the two components can have a small macroscopicity even if they are macroscopic in size (that is, having large energies).

In the following we will consider the macroscopicity of the state 
with respect to the amplitude quadrature, or in other words, we consider the separation of the components of the superposition state in phase space. With this pointer observable, we quantify the macroscopicity with two parameters. The first one is the mean distance between the two macroscopic states in phase space, 
\begin{eqnarray}
D = \frac{1}{\sqrt{2}} |\langle\Phi_+|x|\Phi_+\rangle-\langle\Phi_-|x|\Phi_-\rangle|
\end{eqnarray}
where $x=(a +a^\dagger)/\sqrt{2}$ is the amplitude quadrature. The second quantifying parameter is the success rate in discriminating the two states by a homodyne detector;
\begin{eqnarray}
P=\frac{1}{2}\left(\langle\Phi_+|\Pi_+|\Phi_+\rangle +\langle\Phi_-|\Pi_-|\Phi_-\rangle \right)
\end{eqnarray}
where the measurement projectors have been defined as $\Pi_+=\int_0^{\infty}|x\rangle\langle x|$ for a successful measurement of $|\Phi_+\rangle$ and $\Pi_-=\int_{-\infty}^0|x\rangle\langle x|$ for a successful measurement of $|\Phi_-\rangle$.

Using these phase-space measures for the macroscopicity, the state in Eq. (\ref{jointsub}) does not appear to be macroscopic as its macroscopic components are largely overlapping in phase space. However, it can be simply rewritten as
\begin{equation}      
\ket{\Phi} = \frac{1}{2}\big((\ket{0} + \ket{1}) \ket{\Psi_+} + (\ket{0} - \ket{1}) \ket{\Psi_-}\big)
\label{catRot}
\end{equation}     
where the new macroscopic states are 
$\ket{\Psi_\pm} = \ket{\Phi_+} \pm \ket{\Phi_-} = \sqrt{T} S(r) \ket{0} \pm \sqrt{1-T} \sinh r S(r) \ket{1}$.
In the leftmost column of Fig. \ref{fig:quad} we plot the squared wavefunctions (quadrature probability distributions) of these two macroscopic components for different degrees of squeezing. The solid line wavepackets are associated with states for which we chose $T=1/2$ while the shaded wavepackets correspond to the case where the macroscopic components have equal weights, that is, for $T=\sinh^2 r/(1+\sinh^2 r)$. We see from the plots that the separation, $D$, increases with increasing squeezing, a trend which is quantified in Fig. \ref{fig:mean} by the black lines. The overlap between the states, however, is constant with respect to the squeezing for balanced components (black solid line in Fig. \ref{fig:prob}) while it follows a more complicated structure for $T=1/2$ (black dashed line in Fig. \ref{fig:prob}). Therefore, by choosing a proper splitting ratio, the success in discriminating the two components in a single measurement is quite high (90\%) for any degree of squeezing, but each of the components only become macroscopic for high degrees of squeezing.

\section{Generation of micro-macro states by multi-photon subtraction}
In this section we outline an approach that allows for a further increase of the success rate in discriminating the macroscopic components, $P$, as well as an increase in the phase-space distance, $D$. Instead of using squeezed vacuum as the input to the generation process (Fig. \ref{fig:setups}(c)), we propose to use a photon subtracted squeezed vacuum states. This can be implemented by tapping off a small part of the squeezed beam and registering a certain number of photons using a photon counter (see the optional box in Fig. \ref{fig:setups}(c)). Upon the registration of, say $m$ photons, an $m$-photon subtracted squeezed state is heralded. Experimentally, up to three photons have been subtracted from a squeezed vacuum state~\cite{NIST}. 

\begin{figure}[t]
\begin{center}
\includegraphics[width=\columnwidth]{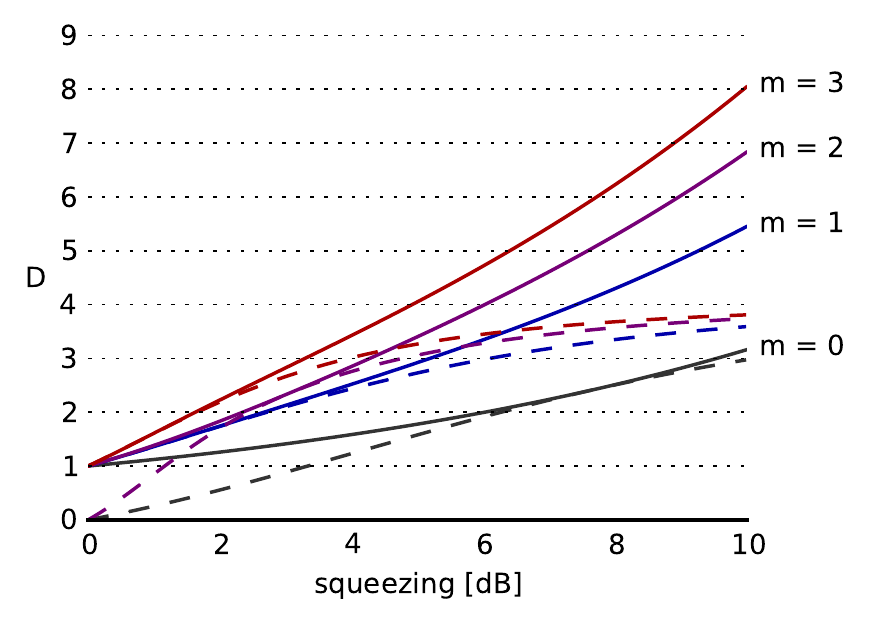}
\caption{(Color online) Mean phase space distance of the two macroscopic states as a function of the degree of squeezing for different photon number subtractions. The dashed curves represent the results for $T=1/2$ while the solid lines are for $T=T_{bal}$.}
\label{fig:mean}

\end{center}
\end{figure}

An m-photon subtracted squeezed state reads~\cite{Jun2012}
\begin{eqnarray}
|\psi^{(m)}\rangle &=&N_m a^m S(r)|0\rangle\\
&=&N_m\frac{1}{\sqrt{\cosh r}}\sum_{k\ge m/2}^\infty\frac{(2k)!(\tanh r)^k}{2^kk!\sqrt{(2k-m)!}}|2k-m\rangle\nonumber
\end{eqnarray}
with normalization $N_m^{-2} = m! (-i\sinh r)^m P_m(i\sinh r)$, where $P_m$ are the Legendre polynomials.  By using this state as the input of the circuit in Fig.~\ref{fig:setups}(c), the output state is as in Eq. (\ref{catRot}),
where the microscopic states are $|0\rangle \pm |1\rangle$ as before, 
but now with
\begin{equation}
\ket{\Psi_\pm} = (\sqrt{T}a^m\pm \sqrt{1-T}a^{m+1})S(r)\ket{0}.
\end{equation}
The two macroscopic components are balanced (i.e. they have equal coefficients after normalization of the state) by setting $T=T_{bal}= \langle n \rangle / (1 + \langle n \rangle)$ where $\langle n \rangle$ is the photon number of the input state. 
To visualize the macroscopicity of the states, we first plot the quadrature probability distributions for $|\Psi_\pm\rangle$ (for the amplitude quadrature) for different photon subtractions $m$ and for different amount of squeezing. The result is illustrated in Fig. \ref{fig:quad} and it is clear that the phase-space distance between the two states increases with increasing squeezing parameter and with increasing number of prior photon subtractions, $m$.
To quantify this effect, in Fig. \ref{fig:mean} we plot the difference of the expectation values, $D$, of the amplitude quadratures of the two states, and it is clear from this plot that the mean distance becomes larger as the squeezing increases. The solid curves represent the balanced case ($T=T_{bal}$) while the dashed curves correspond to $T=1/2$.

% ---------------------------------------------------------
%
%
% ---------------------------------------------------------

% ---------------------------------------------------------
%
\begin{figure}[t]
\begin{center}
\includegraphics[width=\columnwidth]{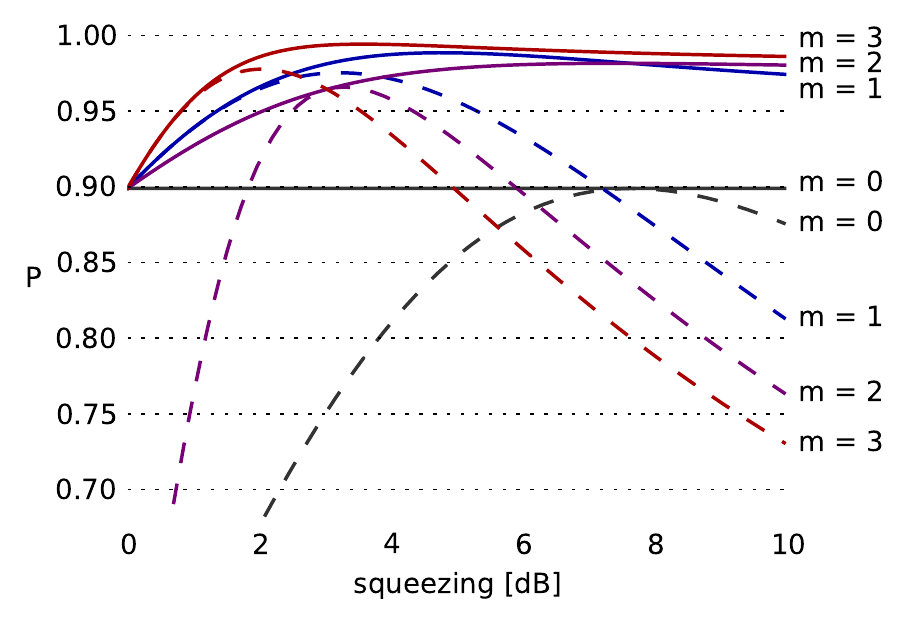}
\caption{(Color online) Success rate in discriminating the two macroscopic states with a dichotomic homodyne detector as a function of the degree of squeezing for different photon number subtractions. The dashed curves represent the results for $T=1/2$ while the solid lines are for $T=T_{bal}$.}
\label{fig:prob}

\end{center}
\end{figure}   
%
% ---------------------------------------------------------

From the plots in Fig. \ref{fig:quad} it seems that the overlap between the states decreases as the squeezing and the number of subtractions increase. However, the exact trend is more complicated as shown in Fig. \ref{fig:prob} where the success rate in discrimination is plotted against the squeezing for a different number of photon subtractions and for $T=T_{bal}$ (solid) and $T=1/2$ (dashed). We see that for low squeezing, the subtraction of an odd number of photons yields a higher success rate than for the subtraction of an even number of photons. However, the general trend (in the case of balanced states) is a rapid increase with the squeezing reaching values well above 90\%. We also note that the rate reaches a maximum for a certain squeezing degree (and slightly decreases for higher values of the squeezing) depending on the photon subtraction number. This behavior is even more pronounced for $T=1/2$. It is caused by the occurrence of side lobes of the wavefunctions for high squeezing degrees which is evident in Fig. \ref{fig:quad} for $T=1/2$ but less clear for $T=T_{bal}$ as in this case the side lobes are very small.           

As the discrimination rate is very close to 100\%, the macroscopicity can be solely described by the mean distance, $D$. Considering for example an input state with a single photon subtracted from a 5~dB squeezed state (corresponding to $m=1$), the distance is $D\approx 3$, corresponding to 6 shot noise units, while the discrimination rate is about 98\%.   

\begin{figure*}[t]
\begin{center}
\includegraphics[width=\textwidth]{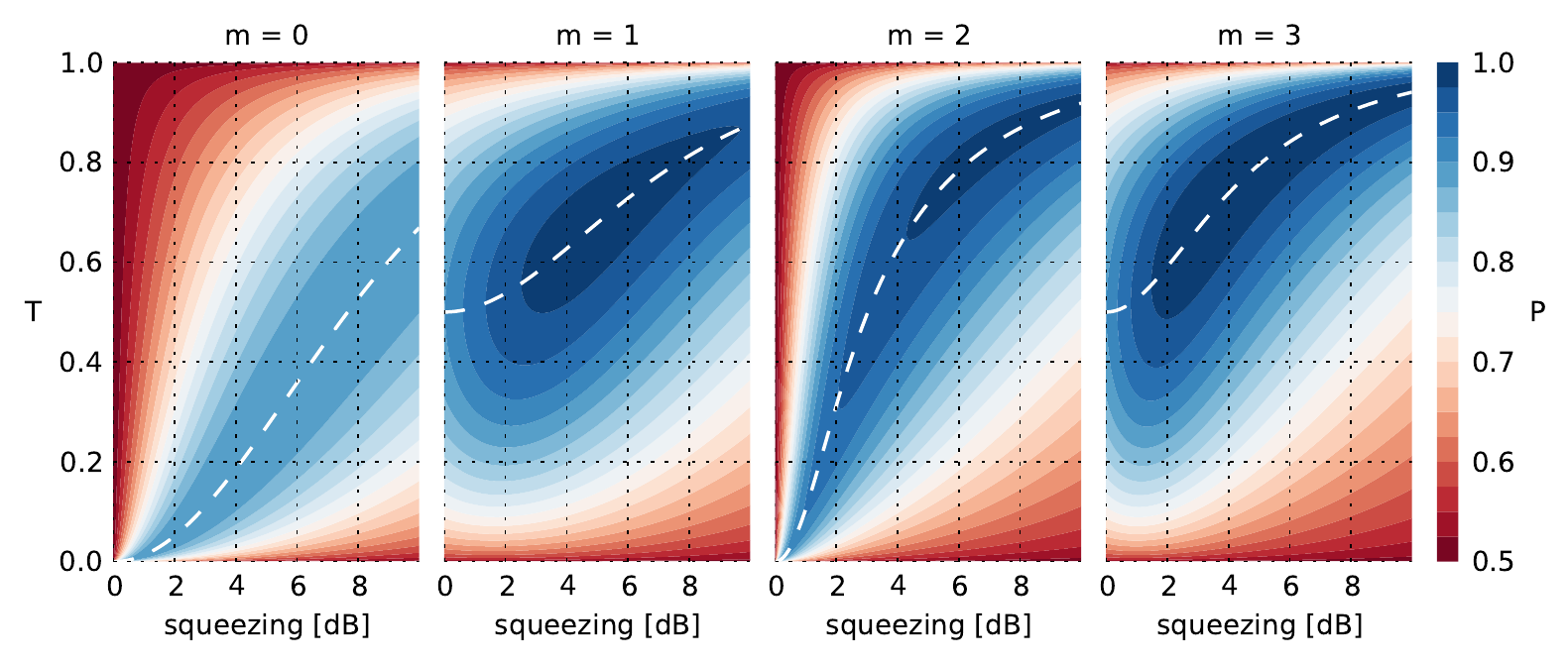}
\caption{Contour plots of the success rate in discriminating the two macroscopic states with a dichotomic homodyne detector as a function of the degree of squeezing and the beamsplitter transmission. The white dashed curves represent the case where the transmission coefficient is set such that the two macroscopic components of the superposition are balanced, $T=T_{bal}$.}
\label{fig:success}

\end{center}
\end{figure*}

Finally, in Fig. \ref{fig:success} we plot the success rate as a function of the transmission $T$ and the squeezing degree. The maximum rate is indicated by the white dashed lines and they correspond to the cases of balanced macro-component for which $T=T_{bal}$.

We note that although we have been considering the macroscopicity with respect to a quadrature measurement, it can be directly translated into a photon number measurement. The states $|\Psi_-\rangle$ and $|\Psi_+\rangle$ cannot be discriminated by an intensity measurement as the information about the two states lies in their phases. However, by performing a simple displacement operation with the amplitude $\beta=D/2$, the state $|\Psi_-\rangle$ will closely resemble the vacuum state while the state $|\Psi_+\rangle$ will be shifted to a higher excited state approximatively containing an average photon number of $|D|^2$. With the example above - single photon subtracted 5dB squeezed state - the two states will after displacement contain approximately 0 and 9 photons, respectively. We note that the success rate in discriminating the two states with a photon detector might be further improved by an additional phase space displacement \cite{modiKen}.

% ---------------------------------------------------------
%
%
% ---------------------------------------------------------

\section{Coherent state superpositions}

The photon-subtracted squeezed states $|\psi^{(m)}\rangle$ are reminiscent of the coherent state superpositions $|\alpha\rangle\pm|\!-\!\alpha\rangle$ \cite{Dakna1997}. However, they are only approximations, and thus to produce the state   
\begin{eqnarray}
|m_+\rangle|\alpha\rangle+|m_-\rangle|\!-\!\alpha\rangle
\label{SCH}
\end{eqnarray}
with macroscopic coherent states, a real coherent state superposition is required at the input to the protocol. By injecting such a state into the generation process (that is, coherently subtracting a single photon from a single-photon state and a coherent state superposition), we find 
\begin{eqnarray}      
\ket{1} \ket{+} &\rightarrow & (\sqrt{T} a_A + \sqrt{1-T} a_B) \ket{1} N_+ (\ket{\alpha} + \ket{-\alpha}) \nonumber\\
&=& \sqrt{T} \ket{0} \ket{+} + \sqrt{1-T} \alpha \frac{N_+}{N_-} \ket{1} \ket{-},
\label{cat2}
\end{eqnarray}
where $|\pm\rangle=N_\pm(|\alpha\rangle\pm|\!-\!\alpha\rangle)$ with normalization constant $N_\pm^{-2} = 2\pm2e^{-2|\alpha|^2}$. This is maximally entangled (balanced) for $\sqrt{T}=\sqrt{1-T}\alpha N_+/N_-$.
However, to obtain the state in (\ref{SCH}) where the macroscopic components are macroscopically distinct in the pointer observable - the amplitude quadrature - we choose the balanced $T$ and rewrite:
\begin{eqnarray}      
(\ref{cat2}) \propto (|0\rangle+|1\rangle)|\alpha\rangle +(|0\rangle-|1\rangle)|\!-\!\alpha\rangle
\label{cat3}
\end{eqnarray}
which is similar to (\ref{SCH}) if the microscopic states are $|m_\pm\rangle=|0\rangle\pm|1\rangle$. We note that it is also possible to transform the state in (\ref{cat2}) into $|0\rangle|\alpha\rangle+|1\rangle|-\alpha\rangle$ using a Hadamard transformation on mode B. The Hadamard transform on coherent state superpositions can be implemented with linear optics and non-Gaussian resources~\cite{Ralph2003,Marek2010}, and has been experimentally demonstrated~\cite{tips}. 

It is clear that for large excitations the two coherent states are macroscopically distinct in their quadrature degree of freedom. The mean phase space separation between the two states is $D=2\sqrt{2}\alpha$ (as also noted in \cite{bjork2004}) and the success rate in discriminating them with the dichotomic homodyne detector is $P=1-(1-\mathrm{erf}(\sqrt{2}\alpha))/2$ \cite{Wittmann2010} which is rapidly increasing with the excitation.

\section{Teleportation}
A microscopic single-photon qubit can be mapped onto a macroscopic qumode using the entangled states proposed in this paper. The entangled state is used in a teleportation protocol with an arbitrary qubit as the input signal: $c_0|0\rangle+c_1|1\rangle$ where $c_0$ and $c_1$ are complex numbers. A Bell measurement that projects onto the four Bell states is jointly performed onto the signal and the entangled state, and the outcome is used to perform a unitary transformation onto the remaining part of the entangled state. A full Bell state measurement is in principle possible~\cite{bjork} but only two projections can be obtained with simple linear optics and vacuum resources \cite{lombardi,lut}. With this scheme it is possible to make the following transformation
\begin{eqnarray}
c_0|0\rangle+c_1|1\rangle\rightarrow c_0|\phi_1\rangle+c_1|\phi_2\rangle 
\end{eqnarray}
where $|\phi_1\rangle$ and $|\phi_2\rangle$ are the macroscopic states $|\Phi_\pm\rangle$, $|\Psi_\pm\rangle$ or $|\pm\alpha\rangle$ depending on which entangled state is being used for the teleportation. Such an operation enables one to link a qubit processor to a qumode processor. 
   
\section{Conclusion}
In conclusion, we have suggested several optical circuits for generating a heralded version of the micro-macro state also known as the optical Schr\"odinger cat state. As opposed to the previous proposals and implementations of a polarization based micro-macro state, the suggested schemes are 
based on heralding which means that the state can be fully characterized with homodyne tomography. The macroscopicity of the micro-macro states has been quantified by two phase-space parameters; the mean distance in phase space between the two macroscopic states and the success rate in discriminating the two states with a homodyne detector. We found that the success rate is above 90\% and that the mean distance is increasing monotonically with the degree of squeezing. Furthermore, it was shown that the state can be used as the resource in a teleporter to map microscopic qubits onto macroscopic qumodes, possibly at remote locations as the entangled state can be efficiently produced at a distance.      

We acknowledge support from the Danish Council for Independent Research (Technology and Production Sciences and Natural Sciences).

{\it Note: Simultaneously with the preparation of this paper, the strategy for preparing a micro-macro state illustrated in Fig.~\ref{fig:setups}(a) was also proposed in~\cite{si}.}

\end{document}